\documentclass[pra,aps,twocolumn]{revtex4-2}
\usepackage{times,epsfig,amssymb,amsfonts,amsmath}

\usepackage{graphicx}
\usepackage{subfigure}
\usepackage{xcolor} 
\usepackage{upgreek}
\usepackage[figuresright]{rotating}
\usepackage{hyperref}

\hypersetup{colorlinks=true,linkcolor=blue,urlcolor=blue,citecolor=blue}

\begin{document}

\title{Long-range bipartite entanglement in XXZ spin chains with the exponential and power-law long-range interactions}

\author{Na Li$^{1,2}$}
\author{Yang Zhao$^{1}$}
\email{zhaoyang22@hebtu.edu.cn}
\author{Wen-Long Ma$^{2,3}$}
\email{wenlongma@semi.ac.cn}
\author{Z. D. Wang$^{4,5}$}
\email{zwang@hku.hk}
\author{Yan-Kui Bai$^{1,4}$}
\email{ykbai@semi.ac.cn}

\affiliation{$^1$College of Physics and Hebei Key Laboratory of Photophysics Research and Application, Hebei Normal University, Shijiazhuang, Hebei 050024, China\\
$^2$State Key Laboratory of Semiconductor Physics and Chip Technologies, Institute of Semiconductors, Chinese Academy of Sciences, Beijing 100083, China\\
$^3$Center of Materials Science and Opto-Electronic Technology, University of Chinese Academy of Sciences, Beijing 100049, China\\
$^4$HK Institute of Quantum Science \& Technology and Department of Physics, The University of Hong Kong, Pokfulam Road, Hong Kong, China\\
$^5$Hong Kong Branch for Quantum Science Center of Guangdong-Hong Kong-Macau Great Bay Area, Shenzhen 518045, China}

\begin{abstract}
Long-range bipartite entanglement (LBE) and its distribution properties are studied in XXZ spin chains with the exponential and power-law long-range interactions (ELRIs and PLRIs). LBE quantified by two-qubit concurrence decays exponentially along with two-site distance in the infinite chain with ELRIs in the thermodynamic limit, and the long-range behavior of two-spin entanglement can detect the quantum phase transition and identify different quantum phases away from the critical point. Moreover, a fine-grained LBE distribution relation is obtained for the infinite XXZ spin chain. On the other hand, in the finite XXZ spin chain with the conventional PLRIs, the long-range concurrence decays algebraically and the total one is no longer monotonic along with the chain length. The total LBE distribution property can exhibit a piecewise function, which has a close relationship with the decaying mode and strength of PLRIs. These LBE relations can be regarded as the generalization of Koashi-Bu\v{z}ek-Imoto bound for the prototypical long-range XXZ model, having potential applications in quantum information processing.
\end{abstract}

\maketitle

\section{Introduction}

Quantum entanglement is a cornerstone for understanding quantum many-body physics \cite{ami08rmp,hor09rmp}, which is not only a physical resource for various tasks of quantum information processing \cite{nie00book,eke91prl,ben93prl,rau01prl,weh18sci,pez18rmp,pan22rmp} but also an effective tool for characterizing quantum properties such as quantum phase transitions (QPTs) and many-body localization  \cite{sac11book,ost02nat,osb02pra,vid03prl,wul04prl,gus04prl,oli06prl,lih08prl,eis10rmp,aba19rmp}. Unlike classical correlations, quantum entanglement cannot be freely shared among many parties \cite{ben96pra,ckw00pra,osb06prl,bai09pra,bai14prl,bai14pra}, and investigating entanglement distribution in many-body systems is a fundamental issue, especially for long-range entanglement \cite{dur05prl,rau05pra,wen10prb,kim13prl,gon17prl,bai22pra,tan24prx,li24njp,bai24prb}. In many-body systems, the long-range bipartite entanglement (LBE) is normally associated with long-range interactions (LRIs) \cite{kof12prl,vod14prl,her17prl}, which is different from the case of short-range interactions \cite{ost02nat,osb02pra,vid03prl}. For example, in an $N$-spin Lipkin-Meshkov-Glick (LMG) model with identical LRIs \cite{lmg65nuc}, there is long-range two-spin entanglement in any spin pair \cite{dus04prl}. But, in a spin-1/2 Ising chain with the nearest-neighbor interactions, two-spin entanglement is short-range and disappears beyond the next-nearest-neighbor spins \cite{ost02nat}.

LRI conventionally, but not universally, refers to the coupling $V$ which decays as a power law along with the distance $r$ between two microscopic subsystems in the large $r$ limit, \textit{i.e.}, power-law LRI (PLRI) $V(r)\sim 1/r^{\alpha}$ with $\alpha$ being the long-range decaying parameter \cite{def23rmp}. In experiments,  LRIs can be precisely engineered and controlled in trapped ions \cite{bri12nat,mai19prl}, Rydberg atoms \cite{luk19sci,gei21sci,sch22prxq}, and superconductor systems \cite{gon21sci}. LRIs have a significant impact on the properties of many-body systems \cite{def23rmp}, such as long-range order in the ground state \cite{pet12prl} and violation of the Lieb-Robinson bound \cite{fre17prb}. However, the LBE distribution, especially under the influence of  LRIs \cite{pet12prl,fre17prb,gon16prb,mag17prl,ren20prb,mon22prb,fra22prb}, is far from clear in multipartite systems. In an $N$-qubit system, although the Koashi-Bu\u{z}ek-Imoto (KBI) bound \cite{koa00pra} and the generalized Coffman-Kundu-Wootters (CKW) monogamy inequality \cite{osb06prl} give the upper bounds on bipartite entanglement distributions, the total two-qubit entanglement is much smaller than these bounds in a realistic long-range multipartite model.

Duseul and Vidal \cite{dus04prl} studied the ground state of an $N$-spin LMG model with identical LRIs and found that a set of bipartite entanglement scaling relations, which provides a more accurate description for LBE distribution in the many-body system. Recently, Xiong \textit{et al} \cite{xio23prb} further extended the study from the finite many-body system to an infinite case by investigating the $p$-wave superconducting model with exponential long-range interactions (ELRI) in the thermodynamic limit, and showed that similar distribution relations are satisfied via the notation of entanglement truncation length. Therefore, it is desirable to explore whether these relations for LBE distribution still hold for other prototypical quantum many-body model. Meanwhile, in comparison to the previous models where LRIs adopt identical or exponential decaying modes, what kind of entanglement distribution relations does the prototypical many-body model obey when LRIs are in the conventional power-law decaying mode?

In this paper, by performing density matrix renormalization group (DMRG) calculations, we study the LBE distribution in the ground state of spin-$\frac{1}{2}$ XXZ chains with ELRIs and PLRIs, respectively. The long-range XXZ spin chain is a prototypical model for investigating quantum magnetism \cite{mat81book}, which can be experimentally engineered in Rydberg atom systems \cite{gei21sci,sch22prxq}. We first consider the infinite XXZ spin chain with ELRIs in the thermodynamic limit, and find that the two-spin entanglement, quantified by the concurrence \cite{woo98prl}, decays exponentially along with two-site distance, and LBE can effectively detect QPT and identify quantum phases. It is further shown that the total bipartite entanglement distribution obeys a fine-grained scaling relations. On the other hand, in the finite XXZ spin chain with the conventional PLRIs, it is found that LBE decays algebraically along with two-spin distance, and a set of new LBE distribution relations are revealed which can exhibit the form of a piecewise function. LBE properties and these new distribution relations in XXZ models have potential applications in quantum information processing.

This paper is organized as follows. In Sec. II, we briefly review the long-range XXZ model, the DMRG algorithm, bipartite entanglement measures, and previously established entanglement distribution relations. In Sec. III, the LBE distribution is studied in the infinite XXZ model with ELRIs, and we show the validity of LBE for the characterization of quantum phases. The entanglement distribution relations for the case of finite XXZ spin chain with conventional PLRIs are studied in Sec. IV. Finally, we give some discussion and conclude our work in Sec. V.

\section{preliminaries}

\subsection{Long-range XXZ model and DMRG algorithm for the ground state}

We focus on the long-range spin-$\frac{1}{2}$ XXZ chain under a transverse field, for which the Hamiltonian has the form
\begin{eqnarray}\label{1}
\mathcal{H}_{XXZ}&=&\sum_{i<j}f(r_{ij})[J_{xy}(S_{i}^{x}S_{j}^{x}+S_{i}^{y}S_{j}^{y})-J_{z}S_i^zS_j^z]\nonumber\\
&&+h_x\sum_{i}S_{i}^{x},
\end{eqnarray}
where the indices $i$ and $j$ run over all sites of the spin chain with open boundary condition, the function $f(r_{ij})=\mbox{exp}[-\alpha(r_{ij}-1)]$ denotes the ELRI mode and  $f(r_{ij})=1/r_{ij}^\alpha$ represents the PLRI mode, with $r_{ij}=|j-i|$ being two-spin distance and $\alpha$ the long-range decaying strength, $J_{xy}$ and $J_z$ are coupling constants in different directions, $S^{\gamma}=\sigma^\gamma/2$ with $\gamma=x,y,z$ are spin operators on site $i$ or $j$ in which $\sigma^\gamma$ are Pauli matrices, we set $\hbar=1$, and $h_x$ is the strength of transverse magnetic field.

To study the LBE property in the many-body system, we need to calculate the ground state of the long-range XXZ spin model. However, generally, the ground state for the long-range model cannot be obtained analytically. Here, we can get the ground state of the long-range Hamiltonian in Eq. \eqref{1} by performing the DMRG calculations \cite{whi92prl,ver04prl,sch05rmp,gro08prb,pir10njp,sch11ap}. The DMRG algorithm is a powerful and preferable numerical tool to obtain the exact ground-state eigenvector of low-dimensional spin models \cite{zhao12prb,zhao23pra}. In this paper, we use the popular TeNPy Library \cite{tenpy18spln} to carry out all the DMRG calculations. 

According to the properties of translation invariance for matrix product operators (MPOs) associated with the many-body model, we use the infinite-size DMRG (iDMRG) method to calculate the ground state of the infinite XXZ spin chain with ELRIs and the finite-size DMRG (fDMRG) algorithm for the case of the finite XXZ spin chain with PLRIs. In all DMRG calculations, we choose the cut of matrix dimension as large as 500 with the truncation error $\epsilon<10^{-10}$, use about $200$ sweeps (the number of cycles to update the wave function) to obtain the ground state of the many-body system with a high degree of precision, and the relative error of the ground state energy is $<10^{-10}$. Moreover, in the case of finite XXZ spin chain with the PLRIs, we choose different chain lengths from 50 to 150, where it is noted that the boundary effect is inevitable in fDMRG calculations, especially for the open boundary condition \cite{shi11prb,hot13prb}. Therefore, in our calculations, we will carefully choose entangled sites by discarding boundary ones to ensure computational accuracy \cite{hua20prb}. Other technical details on the DMRG algorithm are presented in Appendix A.

\subsection{Bipartite entanglement measures and entanglement distribution relations}

For a two-spin reduced state $\rho_{ij}$ with two-site distance $d=|j-i|$ in the long-range XXZ chain, its bipartite entanglement can be quantified by the measure of two-qubit concurrence, which can be expressed as \cite{woo98prl}
\begin{equation}\label{2}
    C_d(\rho_{ij})=\mbox{max}\{0,\sqrt{\lambda_{1}}-\sqrt{\lambda_{2}}-\sqrt{\lambda_{3}}-
    \sqrt{\lambda_{4}}\},
\end{equation}
where we use the translation invariance of the Hamiltonian in Eq. \eqref{1}, and $\{\lambda_i\}$ in descending order are the eigenvalues of the matrix $R=\rho_{d}(\sigma^{y}\otimes\sigma^{y})\rho^{*}_{d}(\sigma^{y}\otimes\sigma^{y})$. For an $N$-qubit state with any pair of qubits being entangled, it is shown that the two-qubit concurrence obeys $C_d(\rho_{ij})\leq 2/N$, which is known as the KBI bound \cite{koa00pra}. Another relevant measure for two-spin entanglement is two-tangle \cite{osb06prl}
\begin{equation}\label{3}
	\tau_d(\rho_{ij})=C_d^2(\rho_{ij}),
\end{equation}
which is the square of two-qubit concurrence \cite{osb05pra}. In an $N$-spin system, the total concurrence and total two-tangle can be defined as 
\begin{equation}\label{4}
	C^{(N)}=\sum_{d=1}^{N-1} C_d, ~~~\tau^{(N)}=\sum_{d=1}^{N-1} \tau_d,
\end{equation}
where $C_d$ and $\tau_d$ quantify two-spin entanglement with the spin-spin distance $d$, and $N$ represents the chain length \cite{note1}. Osborne and Verstraete \cite{osb06prl} proved that the generalized CKW inequality $\sum \tau_d(\rho_{1j})\leq \tau_{1|23\cdots N}$ is satisfied for an arbitrary $N$-qubit state, in which $\tau_d(\rho_{1j})$ is the two-tangle between the central qubit $1$ and qubit $j$, with $j\in \{2,3,\cdots,N\}$, and $\tau_{1|23\cdots N}$ is the linear entropy of reduced density matrix $\rho_{1}$ quantifying bipartite entanglement between the central qubit $1$ and residual qubits.

To accurately characterize the entanglement distribution in a realistic many-body model, Dusuel and Vidal \cite{dus04prl} studied an $N$-spin LMG model with identical LRIs between any two spins and found the entanglement scaling relations
\begin{equation}\label{5}
	C^{(N)}\sim N \tau^{(N)},~~~\tau^{(N)}\sim C_d(\rho_{ij})\sim \frac{1}{N},
\end{equation}
where $C^{(N)}$ and $\tau^{(N)}$ are the total concurrence and total two-tangle defined in Eq. \eqref{4}, and the two-spin concurrence has the property $C_d(\rho_{ij})=C_1(\rho_{i,i+1})$ due to the symmetry. 

Recently, Xiong \textit{et al.} \cite{xio23prb} further extended the study of the LBE distribution to an infinite many-body system and investigated entanglement scaling relations in the $p$-wave superconducting model with exponentially decaying LRIs in the thermodynamic limit. By virtue of the notation of entanglement truncation length $\xi$ (the entanglement quickly vanished beyond this spin-spin distance), they showed that the LBE distributions in the infinite system obey the relations 
\begin{equation}\label{6}
	C^{(\infty)}\sim \xi\tau^{(\infty)},~~~\tau^{(\infty)}\sim C_1(\rho_{i,i+1})\sim \frac{1}{\xi},
\end{equation}
where $C^{(\infty)}=\sum_{d=1}^{\infty}C_d=\sum_{d=1}^{\xi}C_d$ and $\tau^{(\infty)}=\sum_{d=1}^{\infty}\tau_d=\sum_{d=1}^{\xi}\tau_d$
are the total bipartite entanglements, with $\xi$ being entanglement truncation length, and  $C_1(\rho_{i,i+1})$ is the nearest-neighbor two-spin concurrence.

In the following two sections, by performing DMRG calculations, we will present a comprehensive study on LBE distributions in long-range XXZ spin chains with ELRIs and PLRIs, respectively.

\section{LBE distribution in infinite XXZ spin chain with ELRIs}

\begin{figure}
	\epsfig{figure=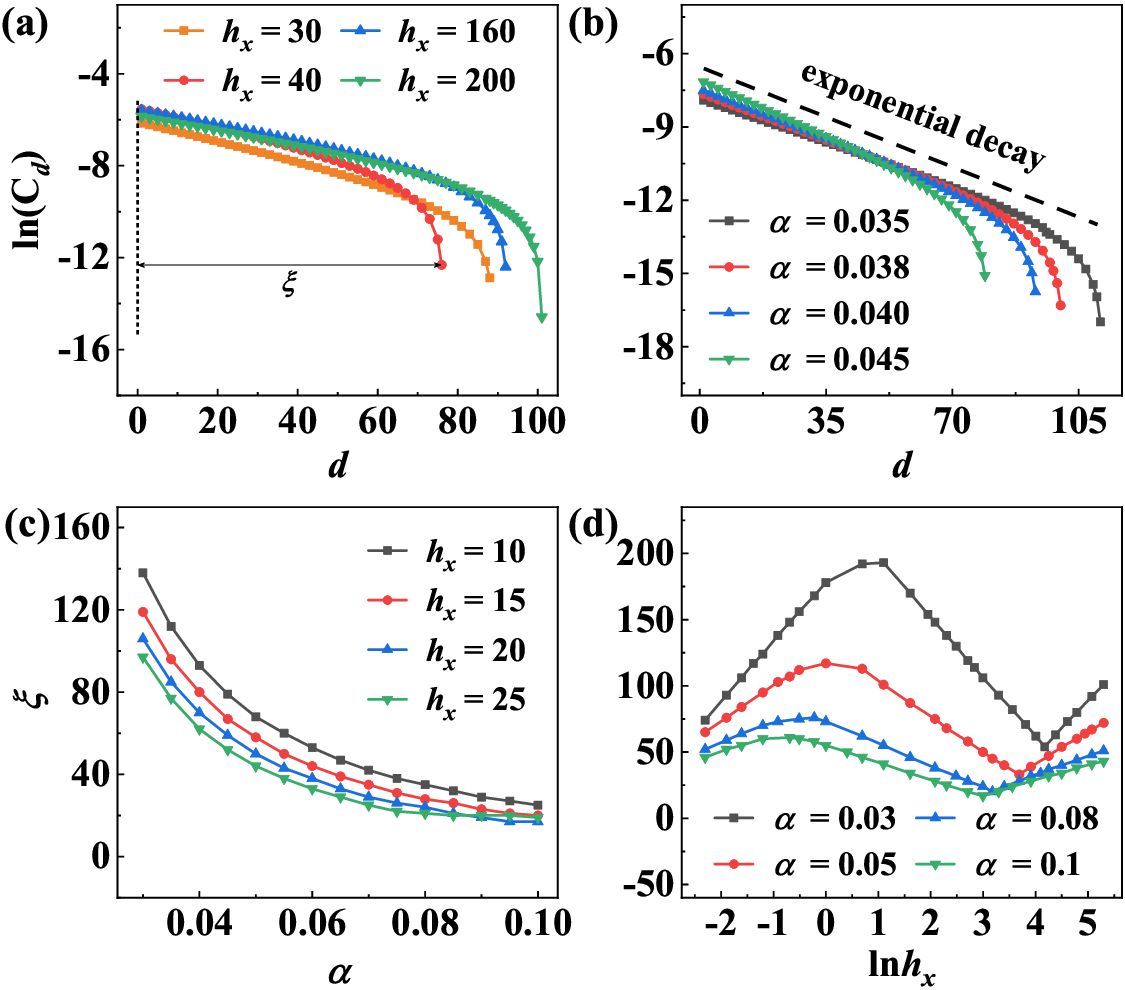,width=0.48\textwidth}
	\caption{LBE in the XXZ spin chains with ELRIs. (a) and (b) Two-spin concurrence $C_d$ as a function of two-site distance $d$ with different values of transverse field $h_x$ and long-range parameter $\alpha$, where $\xi$ is entanglement truncation length, and $C_d$ decays exponentially with $\alpha=0.03$ in (a) and $h_x=10$ in (b). (c) and (d) Effects of $\alpha$ and $h_x$ on the truncation length $\xi$, which are two dominant factors to modulate the long-distance two-spin entanglement.}
	\label{Fig1}
\end{figure}

\begin{figure}[b]
	\epsfig{figure=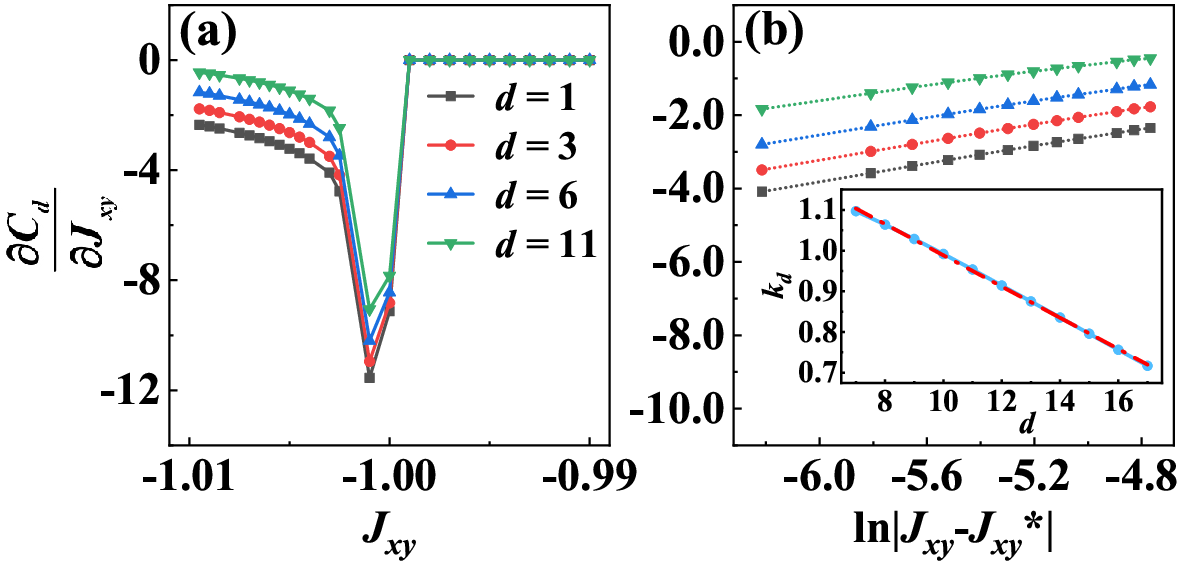,width=0.48\textwidth}
	\caption{Scaling behaviors of two-spin concurrence induced by LRIs. (a) The first derivative of $C_d$ with respect to the coupling strength $J_{xy}$, where the critical point is $J_{xy}^{\ast}=-1.001$. (b) Scaling properties of $\partial C_d/\partial J_{xy}$, and the inset shows a linear function of $k_d$ with respect to two-spin distance $d$ with the parameters $m=-0.3838$ and $m'=1.3724$ in Eq. (\ref{7}).}
	\label{Fig2}
\end{figure}

\begin{figure*}
	\centering
	\includegraphics[width=0.88\textwidth]{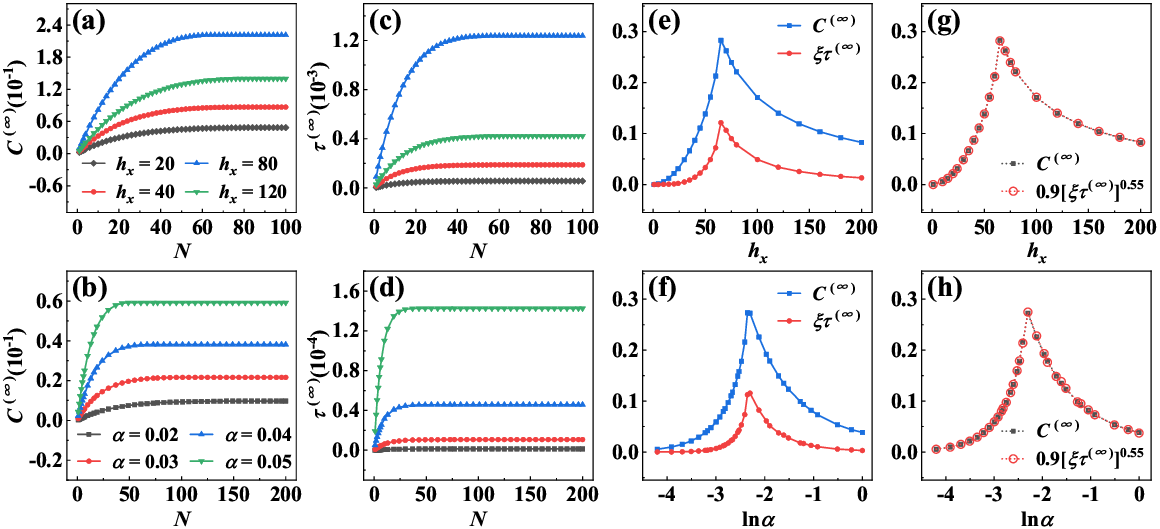}
	\caption{LBE distributions in the infinite XXZ chain with ELRIs. (a) and (b) Total two-spin concurrence $C^{(\infty)}$ as a function of particle number $N$ with the fixed parameters $\alpha=0.03$ in (a) and $h_x=20$ in (b). (c) and (d) Total two-tangle $\tau^{(\infty)}$ as a function of $N$ corresponding to the same parameters in (a) and (b). (e) and (f) Qualitative relationship between $C^{(\infty)}$ and $\xi\tau^{(\infty)}$ with the fixed parameters $\alpha=0.03$ in (e) and $h_x=20$ in (f). (g) and (h) Corresponding fine-grained relation between $C^{(\infty)}$ and $\xi\tau^{(\infty)}$ in Eq. (\ref{8}) with the same fixed parameters.}
	\label{Fig3}
\end{figure*}

Now we study the infinite XXZ spin chain with ELRIs, for which the long-range function in the Hamiltonian of Eq. (\ref{1}) has the exponentially decaying mode $f(r_{ij})=\mbox{exp}[-\alpha(r_{ij}-1)]$ and the coupling constants are chosen to be the typical values $J_{xy}=1$ and $J_z=1$, respectively. As shown in Figs. \ref{Fig1}(a) and \ref{Fig1}(b), we plot the two-spin entanglement $C_d$ as a function of two-site distance $d=|j-i|$ with different values of the decaying strength $\alpha$ and the transverse field $h_x$. It is found that the LBE $C_d$ decreases exponentially in the two figures. Similar to the infinite system in Ref. \cite{xio23prb}, entanglement truncation length $\xi$ is introduced to denote the range of the entanglement $C_d$ since ELRIs only entangle spins in a finite range. Beyond the distance $\xi$, the two-spin quantum state is separable, and the entanglement $C_d$ with $d>\xi$ quickly tends to zero, as shown in Figs. \ref{Fig1}(a) and \ref{Fig1}(b). To determine the length $\xi$, we set a precision $10^{-9}$ for the entanglement $C_d$. In Figs. \ref{Fig1}(c) and \ref{Fig1}(d), we further study the change of truncation length $\xi$ as a function of the long-range parameter $\alpha$ and the transverse field $h_x$, respectively. Given a fixed value of magnetic field $h_x$, as indicated in Fig. \ref{Fig1}(c), we find that the truncation length $\xi$ diverges as the long-range parameter $\alpha\rightarrow 0$ and decreases along with the increasing of $\alpha$, which illustrates that $\xi$ grows along with the range of ELRIs. Interestingly, given a fixed value of the decaying strength $\alpha$, as indicated in Fig. \ref{Fig1}(d), we find that the truncation length $\xi$ is not monotonic along with the increasing of magnetic field $h_x$, and the length $\xi$ can reach a relatively greater value in a weaker magnetic field. This phenomenon of the entanglement is due to the competition between spin-spin coupling and the transverse field in the Hamiltonian. When the magnetic field is small and $h_x\ll J_z=J_{xy}=1$, entanglement in the ground state of the XXZ chain near the ferromagnetic-paramagnetic phase transition has a shorter truncation length. Along with the increasing of $h_x$, the local spin magnetic moments tend to be polarized along the $x$ axis, which exerts influence on the property of the ground state and induces a round peak of the truncation length $\xi$. Therefore, in the XXZ model with ELRIs, both the long-range parameter $\alpha$ and the magnetic field $h_x$ are dominant factors to induce long-distance two-spin entanglement.

In addition, LBE can serve as an effective indicator for QPTs \cite{bai22pra,bai24prb,xio23prb}. Here we use the first derivative of long-range two-spin entanglement $\partial C_d/\partial J_{xy}$ to detect QPT in the infinite XXZ spin chain, where a paramagnetic-ferromagnetic phase transition occurs at the point $J_{xy}/J_{z}=-1$ in the absence of magnetic field \cite{fre17prb,gu03pra}. As shown in Fig. \ref{Fig2}(a), the derivatives diverge not only for the nearest-neighbor concurrence $C_1$ but also for the LBE $C_d$ at the same critical point $J_{xy}^{\ast}$. In the calculation, the long-range parameter is chosen to be $\alpha=1$ and the spacial coupling in the $z$-direction is $J_z=1$, and we obtain the critical value $J_{xy}^{\ast}=-1.001$, as shown in Fig. \ref{Fig2}(a) (the difference $10^{-3}$ from the theoretical value $-1$ is due to the limitations of numerical differentiation). Our results in Fig. \ref{Fig2}(b) demonstrate universal scaling behaviors \cite{gu05pra} induced by LRIs
\begin{equation}\label{7}
    \frac{\partial C_d}{\partial J_{xy}}\simeq k_d \mbox{ln}(|J_{xy}-J_{xy}^{\ast}|),~~k_d=m|d|+m',
\end{equation}
where $J_{xy}^{\ast}$ is the critical value of coupling strength, and the slope $k_d$ depends on the value of two-spin distance $d$ in a linear manner [inset of Fig. \ref{Fig2}(b)] with parameters $m=-0.3838$ and $m'=1.3724$. In addition to detecting QPT, the long-range behaviors of bipartite entanglement can identify different quantum phases away from the critical point \cite{bai24prb}. In Fig. \ref{Fig2}(a), given a value of the LRI $J_{xy}$, we find that the derivative  $\partial C_d/\partial J_{xy}$ gradually decays along with the spin-spin distance $d$ in the paramagnetic phase ($J_{xy}<-1$), but the derivative is invariant with the distance $d$ in the ferromagnetic phase ($J_{xy}>-1$) due to the property $C_d=0$ in this phase. Therefore, according to the decay modes of the first derivative of bipartite entanglement along with the spin-spin distance, one can identify the paramagnetic phase (gradually decaying) and the ferromagnetic phase (invariant) away from the critical point.

Next, we study the entanglement distribution relations of the total bipartite entanglements in the infinite XXZ spin chain with ELRIs. For the typical values of coupling constants $J_{xy}=J_{z}=1$, we calculate the total two-spin concurrence $C^{(\infty)}$ and total two-tangle $\tau^{(\infty)}$ in the infinite system with different magnetic field $h_x$ and long-range parameter $\alpha$, where due to the existence of truncation length $\xi$, the total bipartite entanglements $C^{(\infty)}$ and $\tau^{(\infty)}$ are saturated at a finite value of $N$, as shown in Figs. \ref{Fig3}(a)-\ref{Fig3}(b) and \ref{Fig3}(c)-\ref{Fig3}(d). We also compared the changes of two functions $C^{(\infty)}$ and $\xi\tau^{(\infty)}$ along with system parameters $h_x$ and $\alpha$, and the results in Figs. \ref{Fig3}(e) and \ref{Fig3}(f) illustrate that the relation $C^{(\infty)}\sim \xi\tau^{(\infty)}$ in Eq. \eqref{6} is still satisfied qualitatively. However, this qualitative relation is coarse-grained and a more precise relation is what we desire. Based on the qualitative relation, we make a careful analysis on the total two-spin concurrence $C^{(\infty)}$ as a function of $\xi$ and $\tau^{(\infty)}$ in the long-range model, and obtain a fine-grained generalization  of the KBI bound for the XXZ model
\begin{equation}\label{8}
	C^{(\infty)}\sim a\cdot[\xi\tau^{(\infty)}]^{b},~~\tau^{(\infty)}\sim C_1(\rho_{i,i+1})\sim \frac{1}{\xi},
\end{equation}
where the coefficients are $a=0.9$ and $b=0.55$ for the Hamiltonian parameters we chose, and the fine-grained relations are illustrated for the varying magnetic field $h_x$ [Fig. \ref{Fig3}(g)] and the long-range parameter $\alpha$ [Fig. \ref{Fig3}(h)]. The values of two factors $a$ and $b$ depend on the decaying strength of ELRIs and the entanglement truncation length, which has an imitate relation with the long-range parameter $\alpha$ and the transverse field $h_x$. The second relation in Eq. (\ref{8}) is the consequence of exponential decay of two-spin concurrence along with two-site distance [assuming $C_d=C_1\cdot \mbox{exp}(-|d|/\xi)$, see Figs. \ref{Fig1}(a) and \ref{Fig1}(b)]. Then applying the generalized CKW inequality \cite{osb06prl} to the infinite XXZ spin chain system, we have $C_1(\rho_{i,i+1})\sim 1/\xi$ and $\tau^{(\infty)}\sim 1/\xi$ after a qualitative analysis.

We have explored the LBE distribution in the infinite XXZ spin chain with ELRIs, where long-range entanglements exist and can be utilized to indicate QPT and identify different quantum phases. It should be pointed out that the exponential decay of two-spin concurrence $C_d$ is induced by the exponentially decaying LRIs, resulting in the relation $\tau^{(\infty)}\sim C_1\sim 1/\xi$. Interestingly, a fine-grained relation between  total concurrence $C^{(\infty)}$ and total two-tangle $\tau^{(\infty)}$ is obtained, where the values of coefficients $a$ and $b$ depend on system parameters and the decaying mode of LRIs in the XXZ model. The fine-grained relation in Eq. (\ref{8}) is closely related to the decaying mode of LRIs. Therefore, a natural problem arises: whether these relations still hold when LRIs adopt the conventional power-law decaying mode.

\section{LBE distribution in finite XXZ spin chain with PLRIs}

In this section, we explore the properties of LBE distribution in a finite XXZ spin chain with PLRIs, for which the power-law decaying long-range function in the Hamiltonian of Eq. (\ref{1}) has the form $f(r_{ij})=1/r_{ij}^{\alpha}$. Moreover, the spacial coupling constants are the typical values $J_{xy}=1$ and $J_z=1$, respectively. For the finite XXZ spin chain, we require it to be fully linked according to the KBI bounds, which means that an arbitrary two-spin state $\rho_d$ with two-site distance $d=|j-i|$ should be entangled. Here, by modulating the magnetic field $h_x$ and the long-range parameter $\alpha$, we can obtain a fully entangled spin chain.

\begin{figure}
	\epsfig{figure=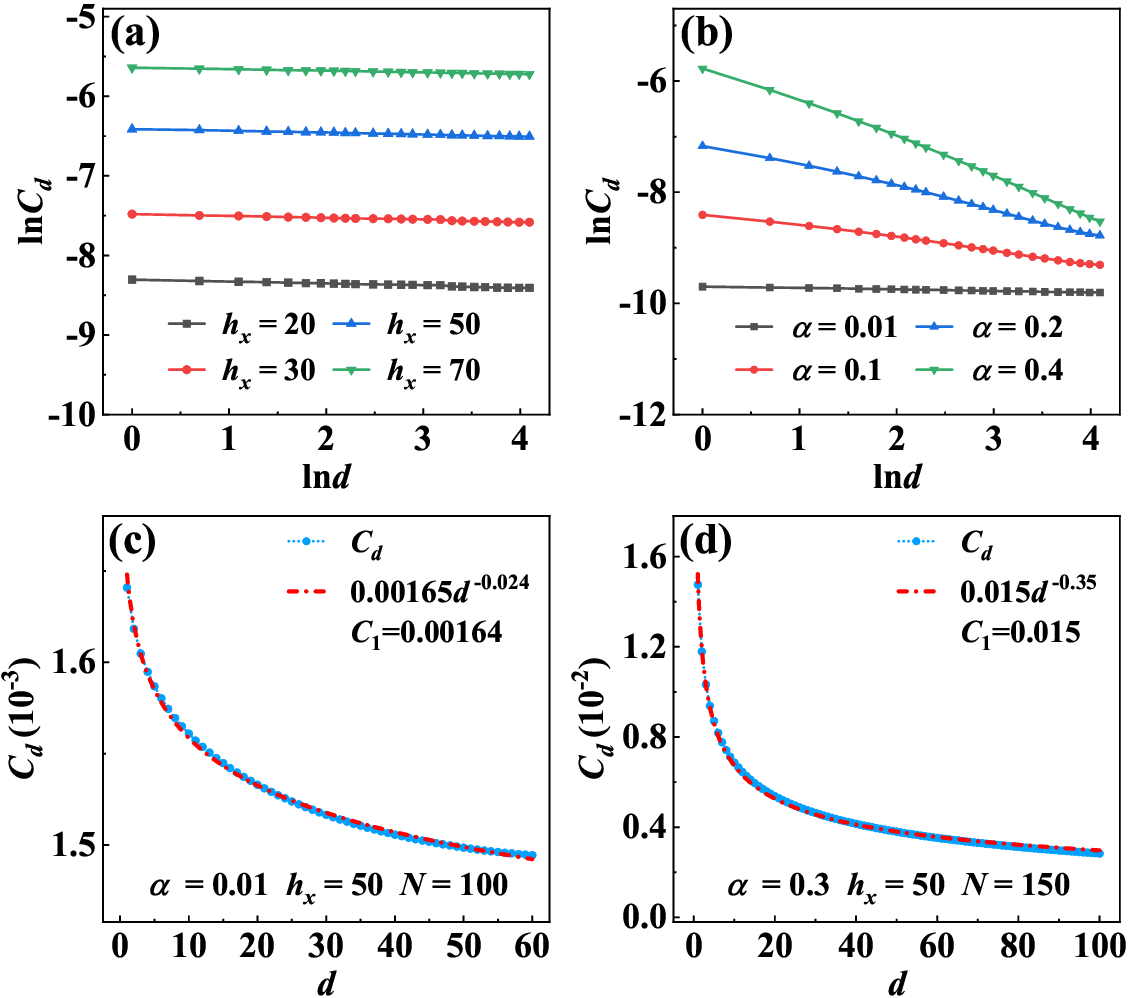, width=0.48\textwidth}
	\caption{Long-range concurrence $C_d$ as a function of two-spin distance $d$ in the finite XXZ chain ($N=100$) with different values of the transverse field $h_x$ and long-range parameter $\alpha$, where the fixed decaying strength $\alpha=0.01$ in (a) and the fixed transverse field $h_x=10$ in (b). Fitting functions of the long-range concurrence $C_d$ for two-site distance $d$ and the nearest-neighbor concurrence $C_1$ [$N=100$ in (c) and $N=150$ in (d)], which exhibit power-law characteristics. Here, due to boundary effects, we discard 40 sites for $N=100$ and 50 sites for $N=150$, respectively.}
	\label{Fig4}
\end{figure}

We first investigate the decaying behavior of long-range two-spin concurrence $C_d$ with respect to two-site distance $d$ in the finite XXZ spin chain, where the translation invariance of the Hamiltonian is considered, and the chain length is chosen to be $N=100$. As shown in Figs. \ref{Fig4}(a) and \ref{Fig4}(b), the two-spin concurrence $C_d$ decays algebraically along with the increasing of two-site distance $d$ for different values of system parameters $\alpha$ and $h_x$. The magnitude of $C_d$ can be modulated by changing the magnetic field $h_x$ and the long-range parameter $\alpha$, and its decay rate is more sensitive to the long-range parameter $\alpha$ [see Fig. \ref{Fig4}(b)]. In addition, we further fit the two-spin concurrence $C_d$ as a function of two-site distance $d$ and the nearest-neighbor concurrence $C_1$ for different system parameters and chain lengths, where it is found that the two-spin entanglement expresses the power-law characteristics $C_d=p\cdot C_1\cdot d^{-q}$ in which the nonnegative values of parameters $p$ and $q$ are related to the chain length, magnetic field and LRIs. As shown in Figs. \ref{Fig4}(c) and \ref{Fig4}(d), two fitting examples for $C_d$ are given, which illustrate that LBEs have power-law decaying behaviors. Here, it should be noted that, due to the boundary effect of finite-size spin chains \cite{shi11prb,hot13prb,hua20prb}, we need to discard some sites near the boundaries to obtain a faithful fitting of the decaying behavior of long-range concurrence $C_d$. Therefore, in the calculations for Fig. \ref{Fig4}, we discard 40 sites for $N=100$ and 50 sites for $N=150$, respectively.

\begin{figure}
	\epsfig{figure=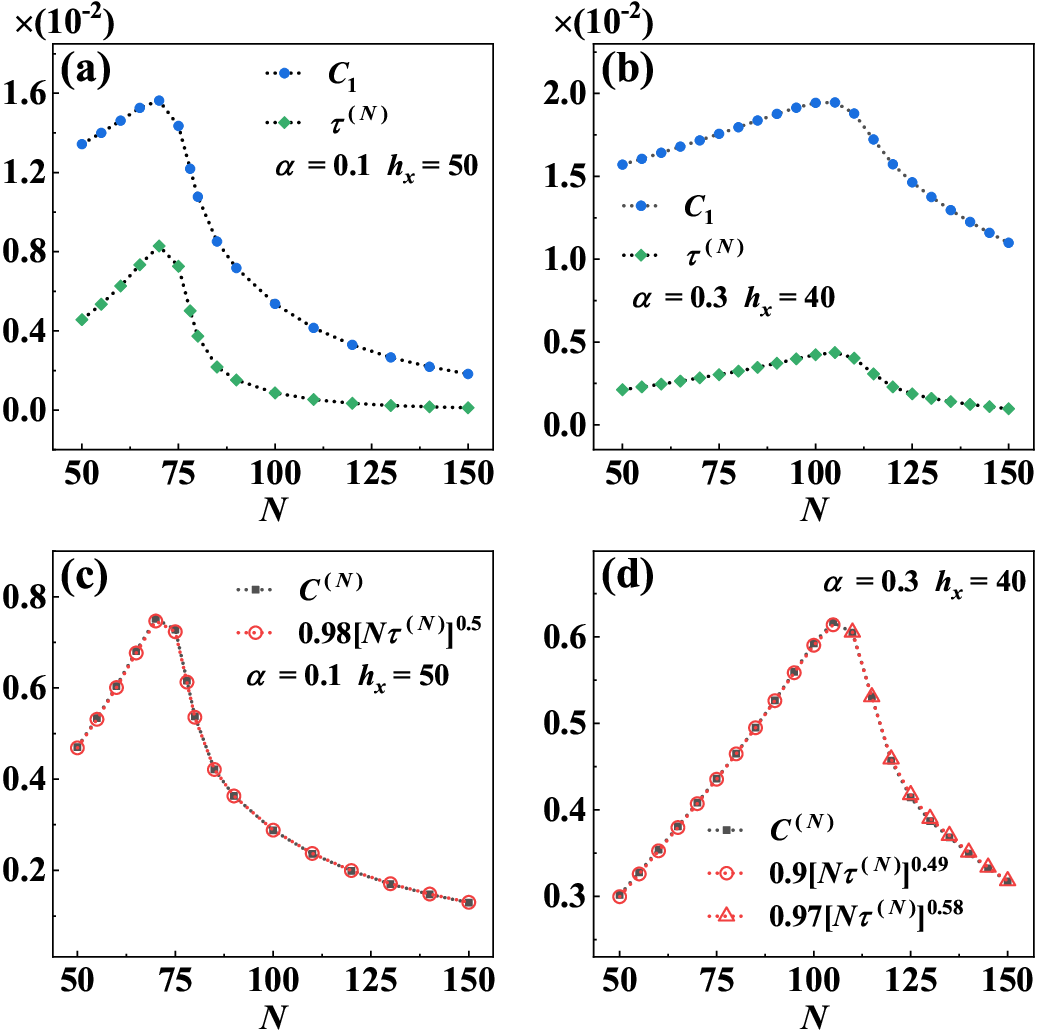,width=0.48\textwidth}
	\caption{LBE distribution in finite XXZ chains with PLRIs. (a) and (b) Proportional relationship between the total two-tangle $\tau^{(N)}$ and the nearest-neighbor concurrence $C_1$, where the two quantities are not monotonic along with the increasing of $N$. (c) and (d) Two kinds of distribution relations between the total bipartite entanglements $C^{(N)}$ and $\tau^{(N)}$ in finite spin chains with different long-range parameter $\alpha$ and transverse field $h_x$.}
	\label{Fig5}
\end{figure}

The distribution of LBE is an important property in many-body systems, and it is desirable to further explore the finite XXZ model with conventional PLRIs. Combineing the power-law decaying long-range concurrence $C_d$ with the monogamy property of two-tangles \cite{osb06prl}, we can obtain
\begin{equation}\label{9}
\tau^{(N)} \sim C_1(\rho_{i,i+1}),
\end{equation}
where the relation implies that the total two-tangle $\tau^{(N)}=\sum_{d=1}^{N-1}\tau_d$ is proportional to the nearest-neighbor concurrence $C_1(\rho_{i,i+1})$. In the derivation, we utilize the property $\tau^{(N)}\sim C_1^2\sum d^{-2q}\sim C_1^2(N-1)\sim C_1$ in which $C_1^2$ is the square of concurrence $C_1(\rho_{i,i+1})$, and we consider the generalized CKW relation with $\tau_d=C_d^2$ \cite{osb06prl} and the KBI bound $C_1\leq 2/N$ \cite{koa00pra} for a fully connected spin chain.

Here, it should be noted that, unlike the second relation in Eq. (\ref{5}) for the finite LMG model with identical LRIs, the total two-tangle $\tau^{(N)}$ and the nearest-neighbor two-spin concurrence $C_1$ are no longer proportional to $1/N$ for the finite XXZ spin chain with the conventional PLRIs. As shown in Figs. \ref{Fig5}(a) and \ref{Fig5}(b), we calculate the entanglement measures $C_1$ and $\tau^{(N)}$ for different system parameters and plot the two quantities as the functions of chain length $N$. It is found that, although the total two-tangle is proportional to the concurrence, \textit{i.e.}, $\tau^{(N)}\sim C_1$, they are not monotonic along with the increasing of chain length $N$. This phenomenon indicates that the LBE distribution in the XXZ spin chain with conventional PLRIs is different from that of the LMG model with identical LRIs, and the main reason results from the different long-range decaying modes of spin-spin interactions. Moreover, we find that the value of nearest-neighbor entanglement $C_1$ of the ground state is determined by the competition between system parameters, and the maximum of $C_1$ depends on the values of the decaying strength $\alpha$ and the transverse field $h_x$, as shown in Figs. \ref{Fig5}(a) and \ref{Fig5}(b). In addition, due to the proportional property between the total two-tangle $\tau^{(N)}$ and the nearest-neighbor concurrence $C_1$ given in Eq. \eqref{9}, we can see that the total two-tangle $\tau^{(N)}$ is not monotonic along with the increasing of chain length $N$ either.

We further study the total bipartite entanglement distribution for $C^{(N)}$ and $\tau^{(N)}$ in the finite XXZ spin chain with PLRIs. After a series of numerical investigations based on the fDMRG algorithm, we find that the KBI bound for the LMG model $C^{(N)}\sim N\tau^{(N)}$ in Eq. \eqref{5} is qualitatively satisfied for the XXZ spin chain when the value of the long-range parameter is small ($\alpha\leq 0.1$). The reason is that, when the decaying strength $\alpha\rightarrow 0$, the power-law decaying parameters $1/r_{ij}^\alpha$s tend to the identical LRIs similar to those of the LMG model. Moreover, following a careful analysis, we obtain a fine-grained relation $C^{(N)}\sim a_1[N\tau^{(N)}]^{b_1}$ for the XXZ model. As shown in Fig. \ref{Fig5}(c), the fitting results are illustrated where the values of $a_1=0.98$ and $b_1=0.5$ are determined by the long-range parameter $\alpha$ and the transverse field $h_x$.

Interestingly, along with the increasing of the long-range decaying strength $\alpha$ for PLRIs, bipartite entanglement distribution will exhibit a piecewise relation 
\begin{equation}\label{10}
C^{(N)}\sim a_1[N\tau^{(N)}]^{b_1}+ a_2[N\tau^{(N)}]^{b_2},
\end{equation}
where $N$ is the chain length, and the coefficients $a_i$ and $b_i$ ($i=1,2$) have close relation with the long-range parameter $\alpha$ and the transverse field $h_x$, and $a_i$'s have the property $a_1=0$ for $N>N_c$ and $a_2=0$ for $N\leq N_c$, with $N_c$ being the key value of chain length corresponding to the maximal total concurrence $C^{(N)}$. In comparison with the LBE distribution in Eq. \eqref{5} for the LMG model, the piecewise expression stems from the decaying mode and strength of PLRIs in the XXZ spin chain which further results in the dependence of the total concurrence $C^{(N)}$ on the chain length $N$. For the Hamiltonian of the XXZ model in Eq. \eqref{1}, the weights of spin-spin interactions and local magnetic field are variable along with the increase in chain length, and the long-range parameter $\alpha$ for PLRIs can further modulate this property, which results in the change of ground state for the multipartite system. Therefore, the total concurrence $C^{(N)}$ is monotonically increasing before $N_c$ and monotonically decreasing after $N_c$, according to different relations for bipartite entanglement distributions. As shown in Fig. \ref{Fig5}(d), we choose the long-range parameter $\alpha=0.3$ for PLRIs and the transverse field $h_x=40$, where the total bipartite entanglements $C^{(N)}$ and $\tau^{(N)}$ exhibit a piecewise relation given in Eq.\eqref{10}, with $a_1=0.9$, $b_1=0.49$, $a_2=0.97$, $b_2=0.58$, and $N_c=107$, respectively.

\begin{table}
  \caption{\label{table} The coefficients $a_i$, $b_i$ and the parameter $N_c$ in Eq. \eqref{10} for the piecewise distribution relations between the total bipartite entanglements $C^{(N)}$ and $\tau^{(N)}$, where the chain length $N$ of the XXZ model with conventional PLRIs is chosen from $50$ to $150$ in the fitting procedure.}
  \begin{center}
   \renewcommand{\arraystretch}{1.5}
   \begin{tabular}{ @{\hskip 0.5cm} c @{\hskip 1.0cm} c @{\hskip 1.0cm} c @{\hskip 0.5cm} }
   \hline\hline
     \mbox{System parameters} & $a_{1}/a_{2}(N_c)$ & $b_{1}/b_{2}$ \\
   \hline
   $\alpha=0.2, h_x=30$ & 0.96/0.98(55) & 0.50/0.52 \\
   $\alpha=0.2, h_x=50$ & 0.96/0.98(98) & 0.50/0.53 \\
   $\alpha=0.3, h_x=20$ & 0.91/0.96(44) & 0.49/0.55 \\
   $\alpha=0.3, h_x=30$ & 0.91/0.96(74) & 0.49/0.57 \\
   $\alpha=0.4, h_x=20$ & 0.80/0.91(59) & 0.46/0.62 \\
   $\alpha=0.4, h_x=30$ & 0.77/0.97(105) & 0.45/0.71 \\
   \hline\hline
  \end{tabular}
  \end{center}
\end{table}

To explain the relation between the coefficients of piecewise relation in Eq. \eqref{10} and the system parameters $[\alpha, h_x]$ of the Hamiltonian in Eq. \eqref{1} with PLRIs, we further calculate LBE and fit the relation of total bipartite entanglement distributions $C^{(N)}$ and $\tau^{(N)}$ for the long-range parameter $\alpha=(0.2, 0.3, 0.4)$, the transverse field $h_x=(20, 30, 50)$, and the chain length $N$ from $50$ to $150$, where the spin-spin interactions are the typical values $J_{xy}=1$ and $J_z=1$, respectively. As shown in Table \ref{table}, the turning point of chain length $N_c$ increases along with the increasing of $\alpha$ and $h_x$, and the exponent $b_1$ is sensitive to parameter $\alpha$, but the exponent $b_2$ is sensitive to both $\alpha$ and $h_x$. Moreover, the coefficients $a_i$ decrease along with the increasing of $\alpha$ and $h_x$, except the case ($\alpha=0.4$ and $h_x=30$), which implies that the values of $a_i$ are the result of joint action of system parameters in the case of a larger value of $\alpha$. 

In this section, we obtain new LBE distribution relations for the finite XXZ spin chain with conventional PLRIs, which can be expressed as the forms given in Eqs. \eqref{9} and \eqref{10}. It is pointed out that the total two-tangle $\tau^{(N)}$ and the nearest neighbor concurrence $C_1$ are no longer proportional to $1/N$, and the total bipartite entanglement distributions between $C^{(N)}$ and $\tau^{(N)}$ can exhibit the form of a piecewise function. These new LBE distribution properties are different from those for the LMG model, which has close relation with the joint action of the power-law decaying mode and the transverse field in the XXZ spin chain.

\section{Discussion and conclusion}

In many-body systems, characterization of LBE is a fundamental issue in quantum entanglement theory since long-distance two-site entanglement is a highly precious resource in quantum information processing. Therefore, it is necessary to explore the LBE property in prototypical long-range XXZ spin chains \cite{mat81book}, beyond the existing the KBI bounds for the finite LMG model \cite{koa00pra,dus04prl} and the infinite $p$-wave superconducting model \cite{xio23prb}. In this work, before analyzing the LBE property in XXZ spin chains, how to generate the long-distance two-spin entanglement is a priority problem. We find that, no matter whether the XXZ spin chain is infinite or finite, both LRIs and external magnetic field are important factors for the generation of  LBE. The joint action of LRIs and the magnetic field can increase entanglement truncation length $\xi$ in the infinite case and generate a fully connected spin chain in the finite case.

In conclusion, we have studied the LBE property and explored its distribution relations in long-range XXZ spin chains with ELRIs and PLRIs. In the infinite spin chain with ELRIs, it is found that the long-range two site concurrence $C_d$ decays exponentially along with the spin-spin distance, and LBE can be utilized to detect the paramagnetic-ferromagnetic phase transition and identify different quantum phases. A fine-grained total bipartite entanglement distribution relation between $C^{(\infty)}$ and $\tau^{(\infty)}$ is obtained via the truncation length $\xi$, as shown in Eq. \eqref{8}. On the other hand, in the finite XXZ spin chain with conventional PLRIs, we show that the long-range concurrence $C_d$ decays algebraically, and the total two-tangle $\tau^{(N)}$ and the nearest-neighbor concurrence $C_1$ are no longer monotonic with respect to $1/N$, as illustrated in Figs. \ref{Fig5}(a) and \ref{Fig5}(b). Interestingly, for the larger long-range parameter $\alpha$, it is revealed that the distribution of the total bipartite entanglements $C^{(N)}$ and $\tau^{(N)}$ can exhibit a piecewise function given in Eq. \eqref{10}. In comparison with the existing relations for the LMG model \cite{dus04prl} and the superconducting model \cite{xio23prb}, these new LBE relations can be regarded as the generalization of the KBI bound for the prototypical long-range XXZ model. In addition to other system parameters in XXZ spin chains, it is pointed out that the decaying modes of LRIs have a close relationship with these new LBE properties. It is still an open problem whether the results in this work are applicable to other typical models, and our preliminary investigations on the long-range XYZ model with ELRIs and PLRIs give a positive answer \cite{note2}.

\section*{Acknowledgments}
This work was supported by NSF-China (Grants No. 11575051, No. 12347165, and No. 12404330), Hebei NSF (Grants No. A2021205020 and No. A2025205002), Hebei 333 Talent Project (No. B20231005), the NSFC/RGC JRS grant (RGC Grant No. N-HKU774/21), the GRF (Grant No. 17303023) of Hong Kong. Y.Z. acknowledges support from Hebei Normal University under Grant No. L2023B06. W.L.M. also acknowledges support from the National Natural Science Foundation of China (Grants No. 12174379 and No. E31Q02BG), the Chinese Academy of Sciences (Grants No. E0SEBB11 and No. E27RBB11), the Innovation Program for Quantum Science and Technology (Grant No. 2021ZD0302300), and the Chinese Academy of Sciences Project for Young Scientists in Basic Research (YSBR-090). The DMRG calculations were performed using the TeNPy Library \cite{tenpy18spln}.

\appendix

\section{Some Details on the DMRG Algorithm} \label{admrg}

In the main text of this article, we obtain the two-spin concurrence of the ground state of the XXZ chain with long-range couplings by performing DMRG calculation. The DMRG algorithm is a famous numerical method to search for the ground state of low-dimensional systems. The essential element of the algorithm is that the physical properties of a subsystem of one model can be faithfully described by the eigenstates of the reduced density matrix of the subsystem \cite{whi92prl,sch05rmp}, where the eigenstates corresponding to the eigenvalues above a threshold $\epsilon$ are kept, and the sum of the discarded eigenvalues is called the truncation error. Therefore, based on the DMRG algorithm, the properties of a large quantum system can be studied in a truncated Hilbert space, avoiding the exponential growth of the Hilbert space with respect to the increasing of the system freedoms.  

To explain the technical details of the DMRG method, we take the XXZ model with ELRIs as an example. As introduced in the main text, the Hamiltonian for this model can be expressed as
\begin{eqnarray}\label{s1}
	\mathcal{H}_{XXZ}&=&\sum_{i<j}e^{-\alpha (r_{ij}-1)}[J_{xy}(S_{i}^{x}S_{j}^{x}+S_{i}^{y}S_{j}^{y})-J_{z}S_i^zS_j^z]\nonumber\\
	&&+h_x\sum_{i}S_{i}^{x},
\end{eqnarray}
where $S_{j}^{x}$ ($S_{j}^{y}$, $S_{j}^{z}$) represents the Cartesian component of spin angular momentum along the $x$ ($y$, $z$) axes on site $j$; $J_{xy}$ and $J_z$ are coupling constants, $r_{ij}=|i-j|$ denotes the distance between site $i$ and site $j$, and $\alpha$ is the decaying strength of LRIs. We choose the eigenstates of $S^z_j$, $|\uparrow\rangle$ and $|\downarrow\rangle$ ($S_j^z|\uparrow\rangle=1/2|\uparrow\rangle$, $S_j^z|\downarrow\rangle=-1/2|\downarrow\rangle$), as the basis vectors on site $j$. $|\uparrow\rangle$ and $|\downarrow\rangle$ define a local Hilbert space $\mathcal{H}_j$ associated with site $j$. Therefore, the Hilbert space $\mathcal{H}$ for a finite chain with length $L$ can be chosen as the direct product of the local Hilbert spaces $\mathcal{H}=\bigotimes_j\mathcal{H}_j$. In such a Hilbert space $\mathcal{H}$, its wave function $|\Psi\rangle$ can be expressed as a linear combination of the basis vectors, such as
\begin{equation}\label{s2}
	|\Psi\rangle=\sum_{\sigma_1\cdots\sigma_L}c_{\sigma_1\cdots\sigma_L}|\sigma_1\cdots\sigma_L\rangle,
\end{equation}
where $\sigma_j=\uparrow$ or $\downarrow$ labels the local basis vectors, the direct product of local basis vectors $|\sigma_1\cdots\sigma_L\rangle$ denotes the basis functions of $\mathcal{H}$, and $c_{\sigma_1\cdots\sigma_L}$ is the weight in the linear combination.

The main aim of the DMRG method is to accurately determine the values of $c_{\sigma_1\cdots\sigma_L}$s in Eq. \eqref{s2} iteratively. Here, we show how to achieve the goal by expressing $c_{\sigma_1\cdots\sigma_L}$ as the product of matrices. By performing the singular value decomposition (SVD), we obtain that the weight $c_{\sigma_1\cdots\sigma_L}$ can be represented by
\begin{equation}\label{MPS}
	c_{\sigma_1\cdots\sigma_L} = M^{\sigma_1}M^{\sigma_2}\cdots M^{\sigma_{L-1}}M^{\sigma_L},  
\end{equation}
where $\{M^{\sigma_j}\}$ denotes a set of matrices labeled by $\sigma_j$ with $j=1,2,\dots,L$, and the elements of matrix $M^{\sigma_j}$ are the variational variables to be fixed by the DMRG algorithm. The maximum dimension of the $L$ matrices is called the bond dimension, which should be properly truncated in DMRG calculation. The wave function $|\Psi\rangle$ is called a matrix product state (MPS) when it is reformulated in the form of Eq. \eqref{MPS}. 

Similar to the procedure of obtaining the MPS in Eq. \eqref{MPS}, the Hamiltonian can also be expressed as the product of matrices, which is termed the matrix product operator (MPO). For instance, $\mathcal{H}_{XXZ}$ has the form
\begin{equation}
	\mathcal{H}_{XXZ} = N^{\sigma_1,\sigma'_1}N^{\sigma_2,\sigma'_2}\cdots N^{\sigma_{L-1},\sigma'_{L-1}}N^{\sigma_L,\sigma'_L},
\end{equation}
where $\{N^{\sigma_j,\sigma'_j}\}$ denotes a set of matrices labeled by two indices $\sigma_j,\sigma'_j$  in contrast with the form of Eq. \eqref{MPS}. To be concrete, for an illustrative case of $L=3$, we have 
\begin{equation}\label{empo1}
	N^{\sigma_1,\sigma'_1} =
	\begin{bmatrix}
		I&S^x&S^y&S^z&h_xS^x
	\end{bmatrix},
\end{equation}
\begin{equation}\label{empo2}
	N^{\sigma_2,\sigma'_2} =
	\begin{bmatrix}
		I  & S^x & S^y & S^z & h_xS^x\\
		0  &   Ie^{-\alpha a} &  0  &   0 & J_{xy}S^x\\
		0  &   0 &  Ie^{-\alpha a}  &   0 & J_{xy}S^y\\
		0  &   0 &  0  &   Ie^{-\alpha a} & -J_{z}S^z\\
		0  &   0 &  0  &   0 & I
	\end{bmatrix},
\end{equation}
\begin{equation}\label{empo3}
	N^{\sigma_3,\sigma'_3} =
	\begin{bmatrix}
		h_xS^x\\
		J_{xy}S^x\\
		J_{xy}S^y\\
		-J_{z}S^z\\
		I
	\end{bmatrix},
\end{equation}
where $I$ is a $2\times 2$ identity matrix; $S^x=\sigma^x/2$, $S^y=\sigma^y/2$, and $S^z=\sigma^z/2$ are components of spin angular momentum, with $\sigma^{x,y,z}$ being the Pauli matrices; and $a$ denotes the lattice constant of the spin chain. 

Then we turn to the main process of the DMRG algorithm. First, prepare an initial wave function $|\Psi\rangle$, which is expressed as a MPS in the form of Eq.~\eqref{MPS}.  Usually, the elements of the matrices $M^\sigma$ with a predefined dimension are set as random numbers. Second, minimize the energy density
\begin{equation}
	E=\frac{\langle\Psi |\mathcal{H}_{XXZ}|\Psi\rangle}{\langle\Psi |\Psi\rangle}.
\end{equation}
To do that, we construct the function $L=\langle\Psi |\mathcal{H}_{xxz}|\Psi\rangle-\lambda\langle\Psi |\Psi\rangle$, with the parameter $\lambda$ being the Lagrangian multiplier. Let $\delta L/\delta \tilde M^{\sigma_j}_{a_j,a_{j+1}}=0$ for each matrix $\tilde M^{\sigma_j}$ (the conjugate of $M^{\sigma_j}$), based on which we can get a group of linear equations with the elements of the matrix $M^{\sigma_j}$ being the variables. Then using the SVD or the Lanczos method, we can find the solution to the system of linear equations. Third, turn to the next matrix $M^{\sigma_{j+1}}$ and repeat the second step. In this way, we can update the matrix $M^{\sigma}$ on each site one by one iteratively and finally stop the iteration when some criteria are satisfied. Generally, we need to do many cycles of updates (sweeps) to achieve a satisfactory convergence. In this paper, we set the number of sweeps as many as $200$ to obtain an accurate ground-energy density with a relative error $<10^{-10}$. More technical details of the DMRG algorithm can be found in Ref.~[\citenum{sch11ap}].

In the main text of this paper, two kinds of DMRG algorithms are employed to calculate the ground states of the long-range XXZ spin chains with ELRIs and PLRIs. The first is iDMRG, which is utilized to deal with models with translation symmetry. For the XXZ model with ELRIs and an infinite chain length $L$, its MPO can be expressed as the form in Eq.~\eqref{empo2} due to the fact that the exponential of a long distance $r_{ij}$ satisfies the property $e^{-\alpha r_{ij}}=e^{-\alpha r_{i,i+1}}e^{-\alpha r_{i+1,i+2}}\cdots e^{-\alpha r_{j-1,j}}$. Therefore, the MPO of $\mathcal{H}_{XXZ}$ and the corresponding MPS are translation invariant. Hence, we can use the iDMRG method to get the ground state of the XXZ model with ELRIs.

The second algorithm is the fDMRG method, which is utilized to deal with models of finite length $L$. In this paper, we use the fDMRG method to find the ground state of the XXZ model with PLRIs, whose MPO cannot be written in a compact translation- invariant form due to the algebraic properties of the power function $1/r_{ij}^{\alpha}$. For the finite XXZ model, we have the Hamiltonian 
\begin{eqnarray}\label{s9}
	\mathcal{H}_{XXZ}&=&\sum_{i<j}1/r^\alpha_{ij}[J_{xy}(S_{i}^{x}S_{j}^{x}+S_{i}^{y}S_{j}^{y})-J_{z}S_i^zS_j^z]\nonumber\\
	&&+h_x\sum_{i}S_{i}^{x}.
\end{eqnarray}
As an example, we consider an illustrative chain length $L=4$ for the finite XXZ model, and its MPO can be expressed as  
\begin{widetext}
	\begin{equation}\label{pmpo1}
		N^{\sigma_1,\sigma'_1} =
		\begin{bmatrix}
			I  & S^x/a^\alpha & S^y/a^\alpha & S^z/a^\alpha & S^x/(2a)^\alpha &\cdots & S^x/(3a)^\alpha & S^y/(3a)^\alpha & S^z/(3a)^\alpha & h_xS^x
		\end{bmatrix},
	\end{equation}
	\begin{equation}\label{pmpo2}
		N^{\sigma_2,\sigma'_2} =
		\begin{bmatrix}
			I  & S^x/a^\alpha & S^y/a^\alpha & S^z/a^\alpha & S^x/(2a)^\alpha & S^y/(2a)^\alpha & S^z/(2a)^\alpha & h_xS^x\\
			0  &   0 &  0  &   0 & 0 &  0  &   0 & J_{xy}S^x\\
			0  &   0 &  0  &   0 & 0 &  0  &   0 & J_{xy}S^y\\
			0  &   0 &  0  &   0 & 0 &  0  &   0 & -J_{z}S^z\\
			0  &   I &  0  &   0 & 0 &  0  &   0 & 0\\
			0  &   0 &  I  &   0 & 0 &  0  &   0 & 0\\
			0  &   0 &  0  &   I & 0 &  0  &   0 & 0\\
			0  &   0 &  0  &   0 & I &  0  &   0 & 0\\
			0  &   0 &  0  &   0 & 0 &  I  &   0 & 0\\
			0  &   0 &  0  &   0 & 0 &  0  &   I & 0\\
			0  &   0 &  0  &   0 & 0 &  0  &   0 & I
		\end{bmatrix},
	\end{equation}
\end{widetext}
\begin{equation}\label{pmpo3}
	N^{\sigma_3,\sigma'_3} =
	\begin{bmatrix}
		I  & S^x/a & S^y/a & S^z/a & h_xS^x\\
		0  &   0 &  0  &   0 & J_{xy}S^x\\
		0  &   0 &  0  &   0 & J_{xy}S^y\\
		0  &   0 &  0  &   0 & -J_{z}S^z\\
		0  &   I &  0  &   0 & 0\\
		0  &   0 &  I  &   0 & 0\\
		0  &   0 &  0  &   I & 0\\
		0  &   0 &  0  &   0 & I
	\end{bmatrix},
\end{equation}
\begin{equation}\label{pmpo4}
	N^{\sigma_4,\sigma'_4} =
	\begin{bmatrix}
		h_xS^x\\
		J_{xy}S^x\\
		J_{xy}S^y\\
		-J_{z}S^z\\
		I
	\end{bmatrix}.
\end{equation}
As shown in Eqs.~\eqref{pmpo2} and \eqref{pmpo3}, these two $N^{\sigma,\sigma'}$s are not translation invariant, and the dimension of matrix $N^{\sigma,\sigma'}$ also grows with $L$.  

Although the fDMRG algorithm can only be employed to deal with finite-size quantum systems, it is powerful in obtaining the ground states of many-body models with the conventional power-law decaying LRIs. During the update process of the DMRG calculation, the local matrices $M^\sigma$ will be optimized back and forth in many cycles, which is termed sweeping. Based on the sweeping process, the ground state wave function can be found accurately with a high degree of precision, even for the two-dimensional finite-size lattices \cite{appd1}.  

In this paper, all DMRG calculations are based on the well-known Python library: TeNPy (Tensor Network Python). We keep the matrix dimension cut $D_c$ as large as $500$ with a truncation error $\epsilon<10^{-10}$. To prove the precision and efficiency of the code, we benchmark our data against the results of the LMG model whose entanglement properties can be evaluated analytically \cite{dus04prl}, finding that our DMRG results coincide exactly with the analytical prediction of the LMG model.


\begin{thebibliography}{99}
\bibitem{note1}   Without loss of generality, we only consider  the total two-spin entanglement between the focused spin and other residual spins.
\bibitem{note2}   We further investigated the long-distance two-site entanglement distribution in the XYZ spin chains with ELRIs and PLRIs, where the coupling constants $J_{xy}=J_z=1$ in Eq. \eqref{1} of the main text are replaced by $J_x=1.2$, $J_y=0.8$ and $J_z=1$. We obtained that the long-range $C_d$ decays exponentially in the infinite XYZ chain with ELRIs and algebraically in the finite case with the conventional PLRIs, and all LBE distribution relations for XXZ chains are still applicable.
\end{thebibliography}
\end{document}